\documentclass[12pt]{iopart}
\expandafter\let\csname equation*\endcsname\relax
\expandafter\let\csname endequation*\endcsname\relax
\usepackage{amssymb,amsbsy,amsmath,amsfonts}
\usepackage{iopams}
\usepackage{diagbox}
\usepackage{braket}
\usepackage{mathrsfs}
\usepackage{graphicx}% Include figure files
\usepackage{epsf,epsfig,float,latexsym,amsthm,fancyhdr,rotating}
\usepackage{graphics,psfrag,longtable}
\usepackage{slashed}
\usepackage{esint}
\usepackage{nicefrac}
\usepackage{braket}
\usepackage{mathtools}

\begin{document}

\title[Electromagnetic Vortex Topologies from Sparse Phased Arrays]{Electromagnetic Vortex Topologies from Sparse Circular Phased Arrays}

\author{H Wang$^1$,
K Szekerczes$^1$\footnote{Present address:
Department of Physics, University of Maryland Baltimore County, Baltimore, MD 21250, USA} and A Afanasev$^1$}

\address{$^1$ Department of Physics, the George Washington University, Washington, DC 20052, USA}

\ead{afanas@gwu.edu}

\begin{abstract}
Structured vortex waves have numerous applications in optics, plasmonics, radio-wave technologies and acoustics. We present a theoretical study of a method for generating vortex states based on coherent superposition of waves from discrete elements of planar phased arrays, given limitations on an element number. Using Jacobi-Anger expansion, we analyze emerging vortex topologies and derive a constraint for the least number of elements needed to generate a vortex with a given leading-order topological charge.
\end{abstract}

%\keywords{structured light, optical vortex, phased array, cat states}
\vspace{2pc}
\maketitle

%\maketitle

\section{Introduction and Motivation}

Excitation of topological states from a discrete number of wave sources was considered for quantum optics with atomic phase arrays \cite{PhysRevLett.119.023603}, plasmonic nano-antenna arrays \cite{Arikawa17,Shutova20}, acoustic actuators \cite{Courtney13}, and RF vortices for quantum networking \cite{Mika20}. Despite a broad variety of above applications, they have one feature in common: They are based on generation of vortex states - either classical or quantum - from a number of discrete elements. The need for {\it sparse} arrays comes from a requirement of keeping the element number low  due to multiple reasons, such as physical limitations in a number of trapped atoms for atomic arrays, nanoscale fabrication capabilities in plasmonics, cost constraints, simplicity and robustness of design, {\it etc}.  

In this study, we address a question of what vortex topologies are possible given a limited number of elements (or wave sources). Previously, the problem was studied in acoustics using numerical simulations \cite{Yang13}.  In optics, formation of vortices from a limited number of plane waves, spherical waves or individual laser beams with controlled relative phases was extensively studied both theoretically and experimentally \cite{Dreischuh02,OHolleran06,Ruben06,Vyas07,Wang09,Xavier12}. 

In this paper, we provide a mathematical formalism for vortex formation with arbitrary element numbers that may be used as a guidance for generation of topological states with sparse phased arrays or a superposition of laser beams. Our formalism is based on Jacobi-Anger expansion (in terms of Bessel functions) that allows us to analyse both leading and sub-leading vortex states and  transition to single-order Bessel vortices as the element numbers increase.  Only linear superposition of waves is considered in this paper. Our motivation for this work is observation and applications of quantum selection rules previously established for optical Bessel beams  Ref.\cite{Afanasev13,Afanasev18}, but for experimental setups with a limited number of wave sources.

\begin{figure}[htbp]
\label{figarray}
  \centering
  \includegraphics[width=5.cm]{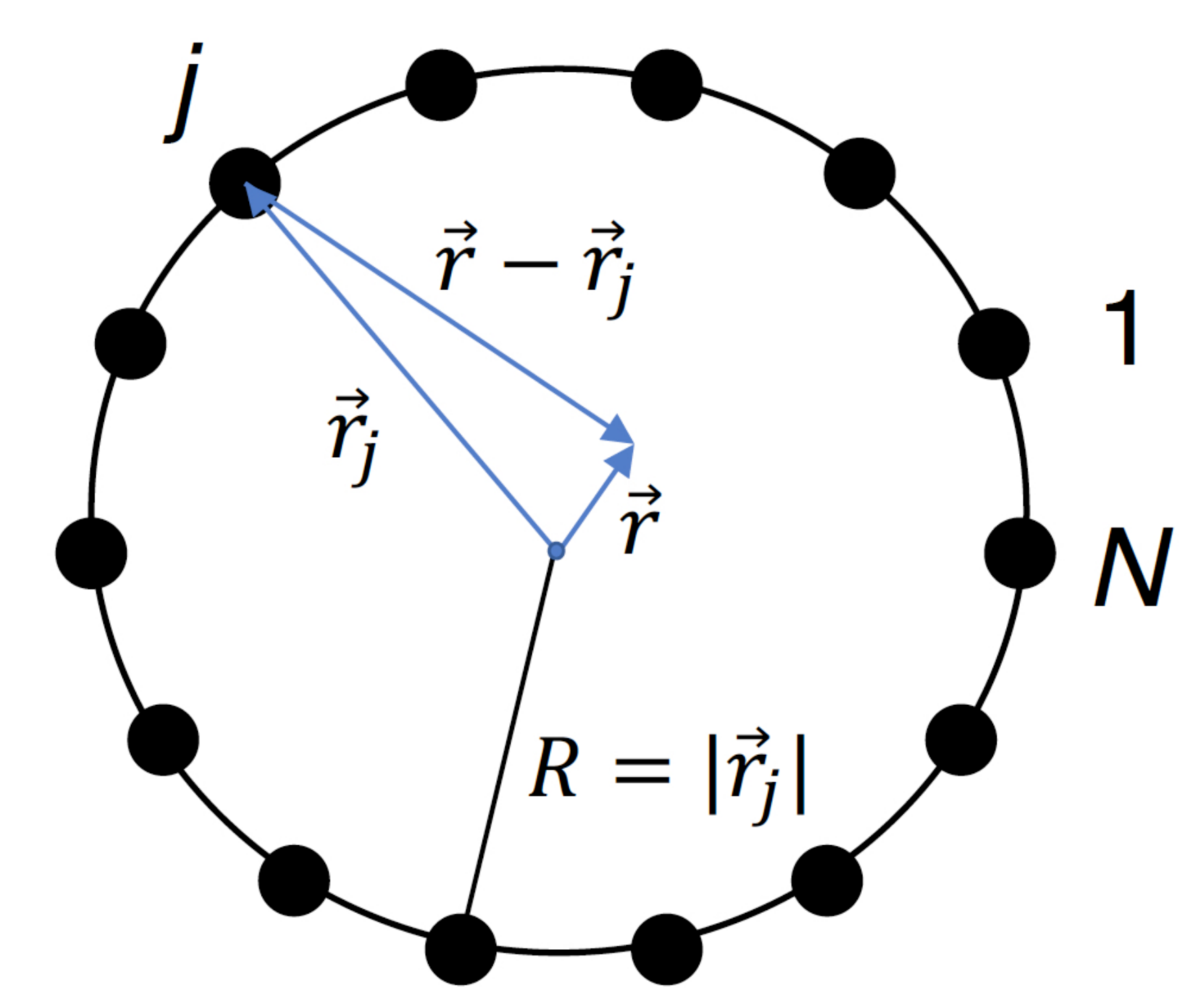}
\caption{A circular phased array of $N$ elements for vortex generation.}
\end{figure}

\section{Formalism for Generation of Vortex States with Phased Arrays}

\subsection{Planar Case}

Consider a circular array of $N$ elements, as shown in Fig.\ref{figarray}, positioned at a radius $R$ and polar angles $\phi_j=2\pi/N$. Each array element generates a spherical scalar wave with an angular-dependent phase $l\phi_j$, where $l$ is an input phase parameter corresponding to the total phase shift of $2\pi l$ around the array and equal phase differences of $2\pi l/N$ between the adjacent elements. The resulting wave amplitude $A(\vec r,t)$ is given by a coherent sum of individual amplitudes,

\begin{equation}
\label{eq:Amps}
    A(\vec r,t)=\sum_{j=1}^{N}
   \frac{A_0\exp{[i(k|\vec r-\vec r_j|+l\phi_j-\omega t)]}}{N|\vec r-\vec r_j|}\equiv A(\vec r)\exp{[-i\omega t]}.
   \end{equation}
Here, $k$ is a wave number and $\omega$ is an angular frequency. Using the approximation  $r\ll R$ in the central area of the array and retaining  terms up to linear in $r$, we have: 
 \begin{equation}
    A(\vec r)\approx\sum_{j=1}^{N}\frac{A_0\exp{[i(kR-kr\cos(\phi-\phi_j)+l\phi_j)]}}{NR} 
\end{equation}

In the limit $N\to\infty$ the above sum is known ($c.f.$ Ref.\cite{Courtney13}),
\begin{equation}
    A(\vec r,t)=\lim_{N\to\infty}\sum_{j=1}^{N}\frac{A_0e^{i(kR-kr\cos(\phi-\phi_j)+l\phi_j-\omega t)}}{NR}=\frac{A_0 e^{ikR}}{R}(-i)^lJ_l(kr)e^{i(l\phi-\omega t)},
\end{equation}
where $J_l(kr)$ is a Bessel function and the polar-angle dependence is given by a factor $e^{il\phi}$, with the entire pattern revolving at an angular speed $\dot{\phi}= \omega/l$ around the origin. Thus, we obtain a Bessel vortex with a topological charge (or a winding number) $l$, see, $e.g.$ Ref.\cite{Courtney13}.  
%We will further omit time dependence for simplicity. 

However, vortex waves may be generated by a smaller number of elements - hence the term {\it sparse arrays} - with an input phase  parameter $l$ not necessarily matching a topological charge of the generated vortex. To analyze the solution for finite $N$, we first use Taylor expansion near the array center $kr\ll 1$ for various combinations of the number of sources $N$ and an input phase parameter $l$. From the result shown in Table~1, we can see that if $l=1$, vortices with a topological charge $n=1$ are generated for $N\geq 3$; while in order to produce $n=2$, at least $N=5$ is required. Interestingly, for $N=2l$, we obtain a superposition of vortices with opposite topological charges, $\ket l$ and $\ket{-l}$: ($\ket 1+\ket{-1}$) for $(N,l)=(2,1)$, ($\ket 2+\ket{-2}$) for $(N,l)=(4,2)$, $etc.$

\begin{table}[htb]
 \centering \caption{First non-vanishing terms of Taylor expansion in $kr$ for an amplitude $A(\vec r)$ formed by an $N$-element phased array with an input phase parameter $l$}
\begin{tabular}{lll}
    \hline
    \diagbox{N}{\textbf{\textit{l}}} &\vline\qquad 1 & 2\\
    \hline
    2 & $-\frac{1}{2R}A_0ikre^{ikR}(e^{i\phi}+e^{-i\phi})$
     & $\frac{1}{R}A_0e^{ikR}$\\
    3 & $-\frac{1}{2R} A_0 ikre^{ikR}e^{i\phi}$ & $-\frac{1}{2R} A_0ikre^{ikR}e^{-i\phi}$\\
    4 & $-\frac{1}{2R} ikre^{ikR}e^{i\phi}$ & $-\frac{1}{8R} A_0(kr)^{2}e^{ikR}(e^{2i\phi}+e^{-2i\phi})$ \\
    5 & $-\frac{1}{2R} A_0ikre^{ikR}e^{i\phi}$ & $-\frac{1}{8R} A_0(kr)^{2}e^{ikR}e^{2i\phi}$ \\
    \hline
   \end{tabular}
    \end{table}
    
To extend our analysis to larger values of $(kr)$ and include contributions of higher-order vortices, we use a Jacobi-Anger expansion \cite{Abramowitz}:
\begin{equation}
\label{eq:JAng}
A(\vec r)=\frac{A_0 e^{ikR}}{NR}\sum_{j=1}^{N}\sum_{n=-\infty}^{\infty}(-i)^nJ_n(kr)e^{in\phi}e^{i(l-n)\phi_j}.
\end{equation}
The summation over the array elements, $\frac{1}{N}\sum_{j=1}^{N}e^{i(l-n)\phi_j}$ can be done recognizing a sum of a geometric series in which a coefficient and a common ratio are equal to $e^{i(l-n)\phi_1}$, with $\phi_1=2\pi/N$:
\begin{equation}
\sum_{j=1}^{N}e^{i(l-n)2\pi j/N}=e^{i(l-n)\frac{2\pi}{N}}\frac{1-e^{i(l-n)2\pi}}{1-e^{i(l-n)\frac{2\pi}{N}}}=\begin{cases}
      N & \text{if } n=l+mN, \: m=0,\pm1,\pm2,...\\
      0 & \text{otherwise}
    \end{cases}  
\end{equation}
Therefore, after summing over the array elements, the remaining sum in Eq.(\ref{eq:JAng}) only includes topological charges $n=l+m N$ of the superimposed vortex states - note that $m$ is now a summation index - and we arrive at the following result:
\begin{equation}
\label{eq:2Dsum}
A(\vec r)=\frac{A_0 e^{ikR}}{R}\sum_{m=-\infty}^{\infty}\left[(-i)^nJ_n(kr)e^{i n\phi}\right]_{n=l+mN}.
\end{equation}
It is the main analytic result of this paper. To the best of our knowledge, in previous analyses of vortex formation from limited sources \cite{Yang13,Dreischuh02,OHolleran06,Ruben06,Vyas07,Wang09,Xavier12} analytic summation over the sources was not performed and Fourier-Bessel representation as in Eq.~(\ref{eq:2Dsum}) was not used.

It can be verified by direct substitution that the above expression Eq.(\ref{eq:2Dsum}) is invariant under transformations $l\to l\pm N$, similar to well-known Born-K\'arm\'an cyclic boundary conditions for a chain of $N$ atoms in crystals.

The series representation of Eq.(\ref{eq:2Dsum}) identifies the orders of Bessel vortices that contribute to a vortex structure as generated by a finite number of elements $N$. The smallest allowed Bessel terms with topological charges $n$ are listed in Table 2 for lowest values of $N$ and $l$. It can be seen that increasing $N$ eliminates all lower-order vortices except $n=l$, which is also the only vortex remaining in $N\to\infty$ limit.  By increasing the element number $N$, we eliminate more harmonics adjacent to $n=l$. The particular choice $N=2l$ results in generation of vortex states with opposite topological charges, i.e., $((-i)^l\ket{l}+(-i)^{-l}\ket{-l})+((-i)^l\ket{3l}+(-i)^{-l}\ket{-3l})...$. Obviously, higher-order topological charges are involved, but choosing a smaller area around the origin where $kr<1$ would lead to power suppression of higher-order states. It follows from Bessel function's behavior at small arguments, $J_n(kr)\propto (kr)^n$, that is also in agreement with leading terms of Taylor expansion from Table 1.

\begin{table}[htb]
 \centering \caption{Lowest allowed topological charges $n$ for Bessel-vortex superposition states produced by an $N$-element phased array with an input phase parameter $l$}
\begin{tabular}{llll}
    \hline
    \diagbox{N}{\textbf{\textit{l}}} &\vline\qquad 1 & 2 & 3 \\
    \hline
    2 &$n=\pm1,\pm3,\pm5,\pm7$
     & $n=0,\pm 2,\pm 4,\pm6$ & $n=\pm1,\pm3,\pm5,\pm7$ \\
    3 &$n=1,-2,4,-5$ & $n=-1,2,-4,5$ & $n=0,\pm3,\pm6,\pm9$ \\
    4 &$n=1,-3,5,-7$ & $n=\pm2,\pm6,\pm10,\pm14$ & $n=-1,3,-5,7$\\
    5 &$n=1,-4,6,-9$ & $n=2,-3,7,-8$ & $n=-2,3,-7,8$\\
    6 &$n=1,-5,7,-11$ & $n=2,-4,8,-10$ & $n=\pm3,\pm9,\pm15,\pm21$\\
    \hline
   \end{tabular}
    \end{table}
    
\subsection{Extension to 3D Case}

Let us extend our approach to three dimensions and consider a wave amplitude of Eq.(\ref{eq:Amps}) in a detection plane located at a distance $z$ away from the circular phased array and oriented parallel to the latter. In cylindrical coordinates, positions of the sources are given by $\vec r_j=(R,\phi_j,z=0)$ and the position in the detection plane is $\vec r=(\rho,\phi,z)$. The distance between an array element $\#j$ and a detection point is written in a far-field (Fraunhofer) approximation,  for large values of $z$, compare with Ref.\cite{Lipson}:
\begin{equation}
    |\vec r_j-\vec r|=\sqrt{R^2+\rho^2+z^2-2R\rho\cos{(\phi-\phi_j)}}\approx z+\frac{R^2}{2z}-\frac{R\rho}{z}\cos{(\phi-\phi_j)}.
\end{equation}

Following the steps of the previous subsection, we obtain the amplitude in such 3D-geometry: 
\begin{equation}
\label{eq:3Dsum}
 A^{3D}(\vec r)=\frac{A_0 e^{ik(z+\frac{R^2}{2z})}}{z}\sum_{m=-\infty}^{\infty}\left[(-i)^nJ_n(k_\perp\rho)e^{i n\phi}\right]_{n=l+mN}\ \ ,
\end{equation} 
where $k_\perp=kR/z$.

We can see that equations (\ref{eq:2Dsum}) and (\ref{eq:3Dsum}) define  same vortex topologies in their respective planes and differ only by overall factors. 

\section{Numerical Examples and Discussion}

In order to demonstrate applications of the above formalism, let us for example take input phase parameters $l$=1 and 2, element numbers $N$=3, 6, and 12, and analyze individual vortex-state contributions to the resulting amplitude near the array's center.

\begin{figure}[htbp]
\label{figarray1}
  \centering
  \includegraphics[width=13.cm]{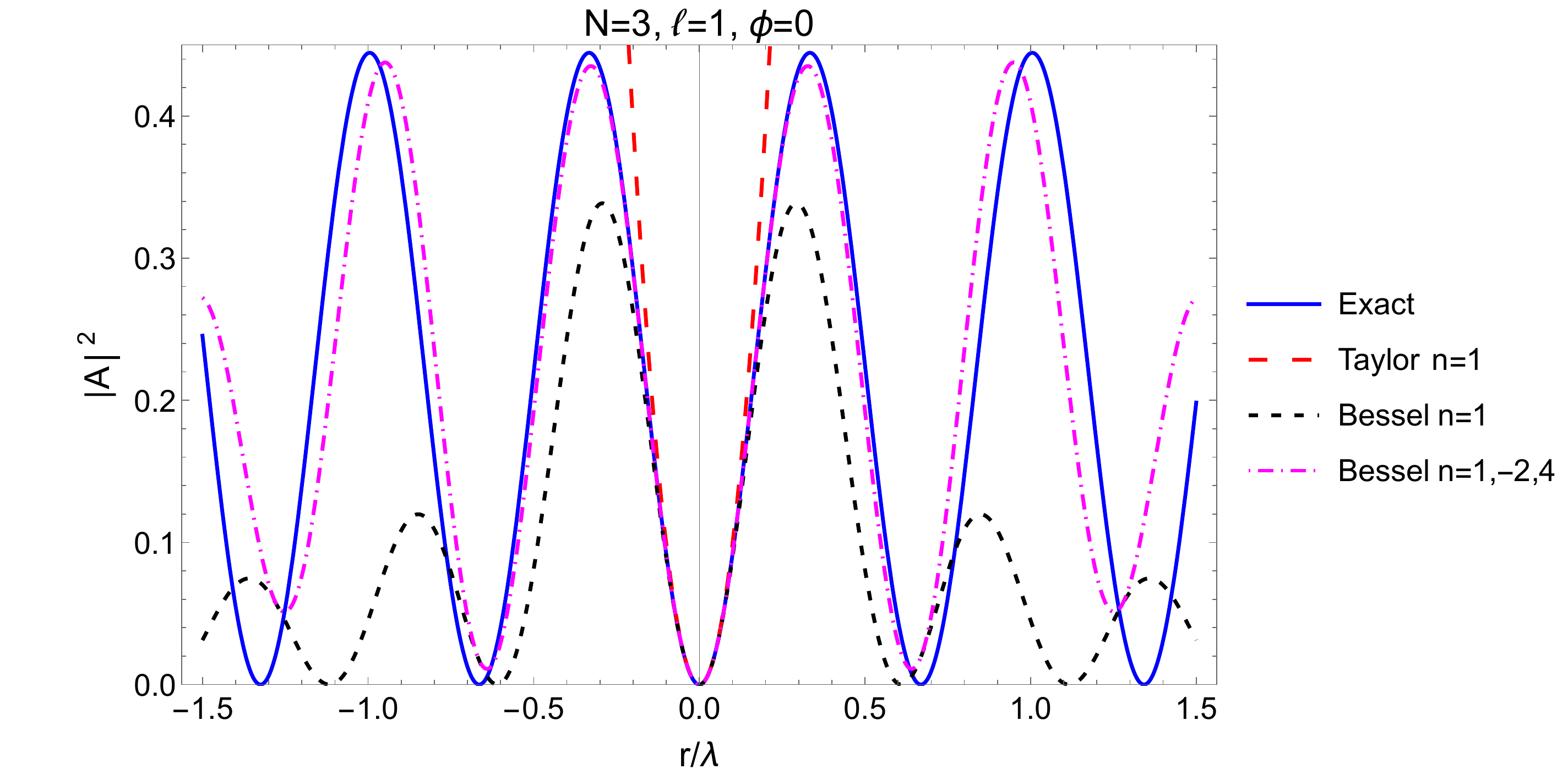}
  \includegraphics[width=13.cm]{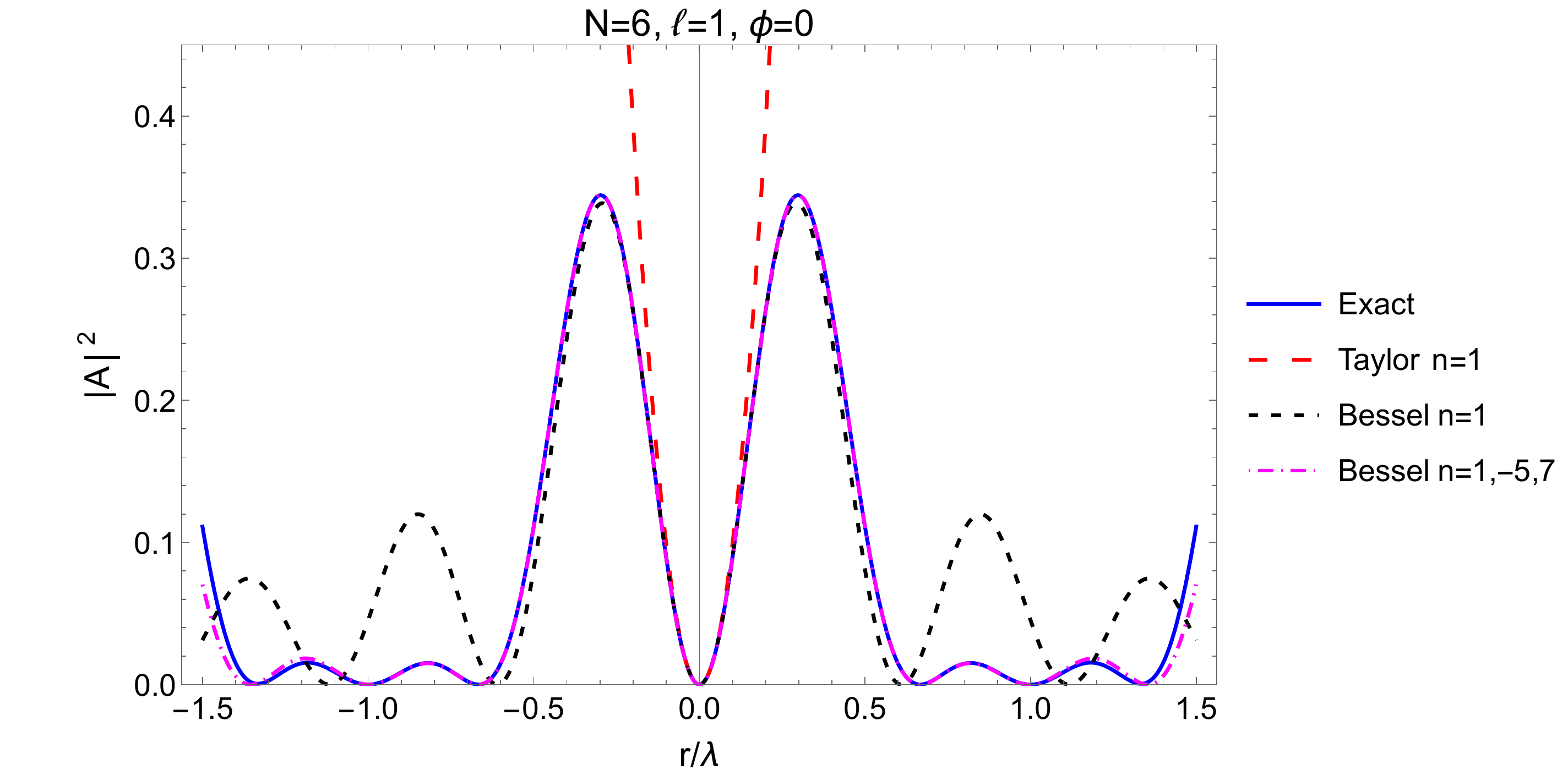}
  \includegraphics[width=13.cm]{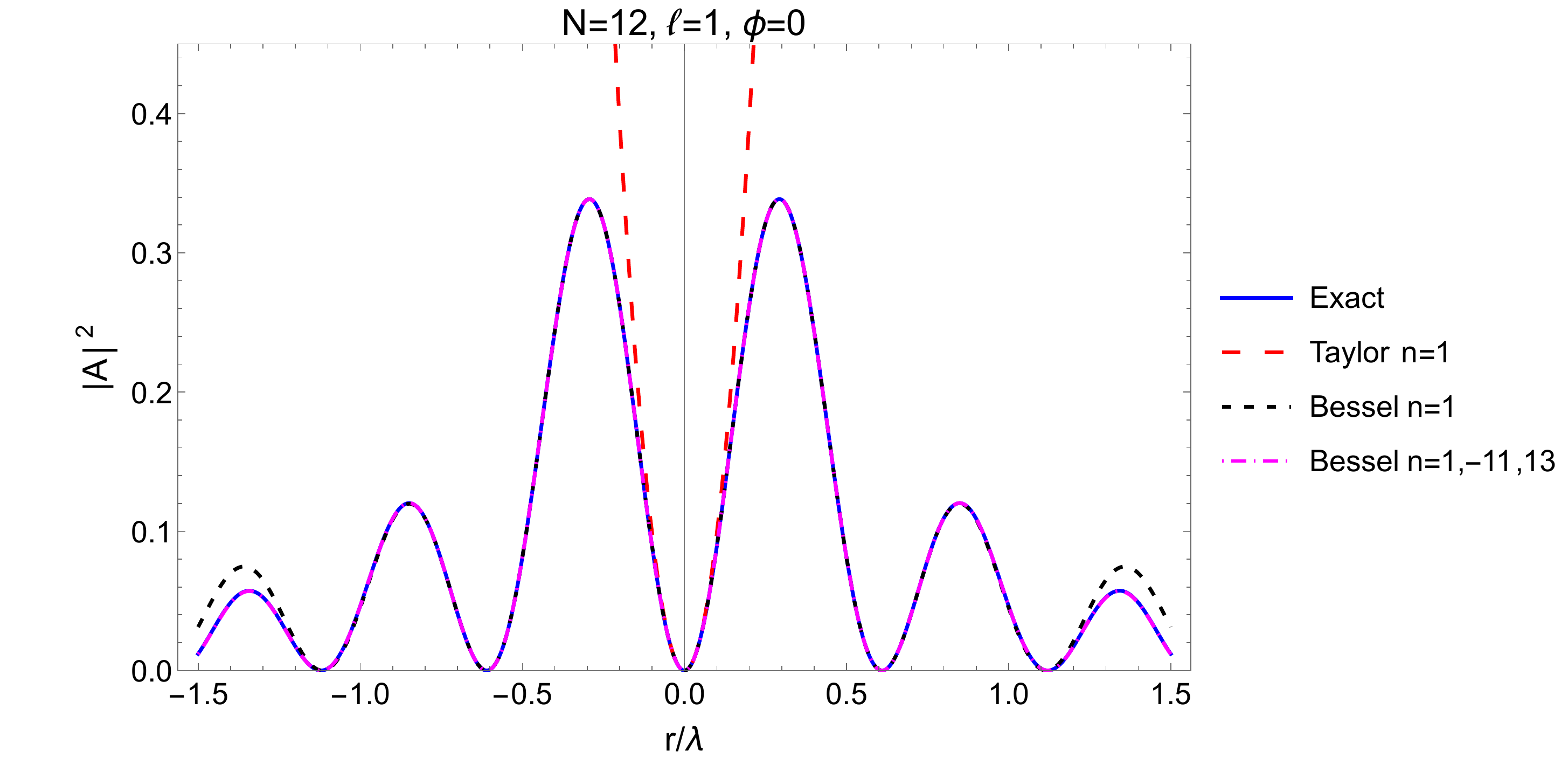}
\caption{Exact vs approximate results  for radial dependence of a time-averaged amplitude squared $|A(\vec r)|^2$ (in arbitrary units) formed by $N$ elements with an input phase parameter $l=1$; $\phi=0$ ($r>0$) and $\phi=\pi$ ($r<0)$ and array radius $R/\lambda$=50. A blue line is a result of Eq.(\ref{eq:Amps}), a red dashed line is the leading term of Taylor expansion in $kr\ll 1$ from Table 1, a black short-dashed line is a lowest-order Bessel term $n=1$ from Table 2 (which also corresponds to $N\to\infty$ limit), and a magenta dash-dotted line includes two additional allowed Bessel terms in expansion of Eq.(\ref{eq:2Dsum}) (see text and Table 2). Element number is (a) $N=3$, (b) $N=6$ and (c) $N=12$. }
\end{figure}

\begin{figure}[htbp]
\label{figarray2}
  \centering
  \includegraphics[width=13.cm]{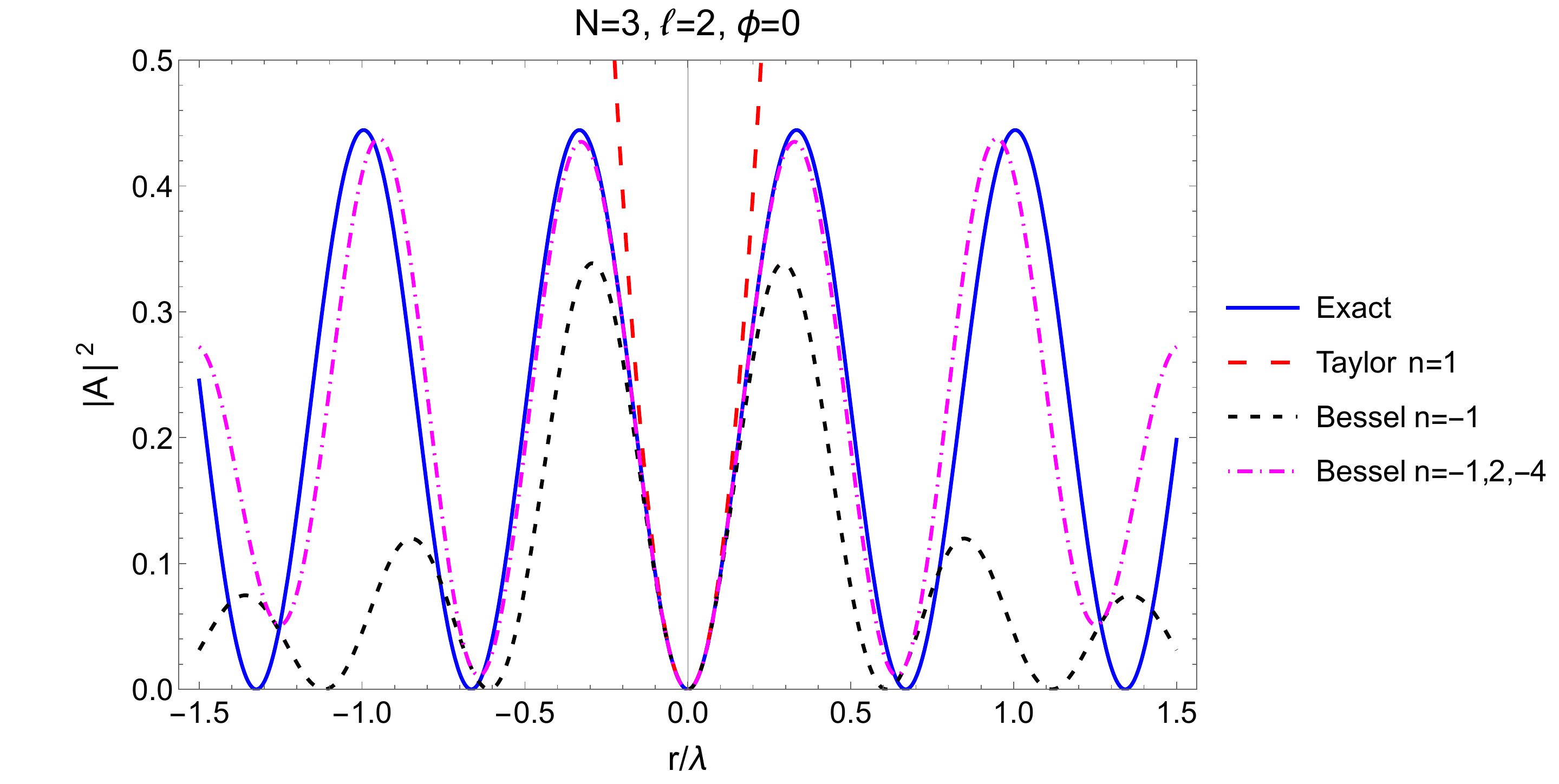}
  \includegraphics[width=13.cm]{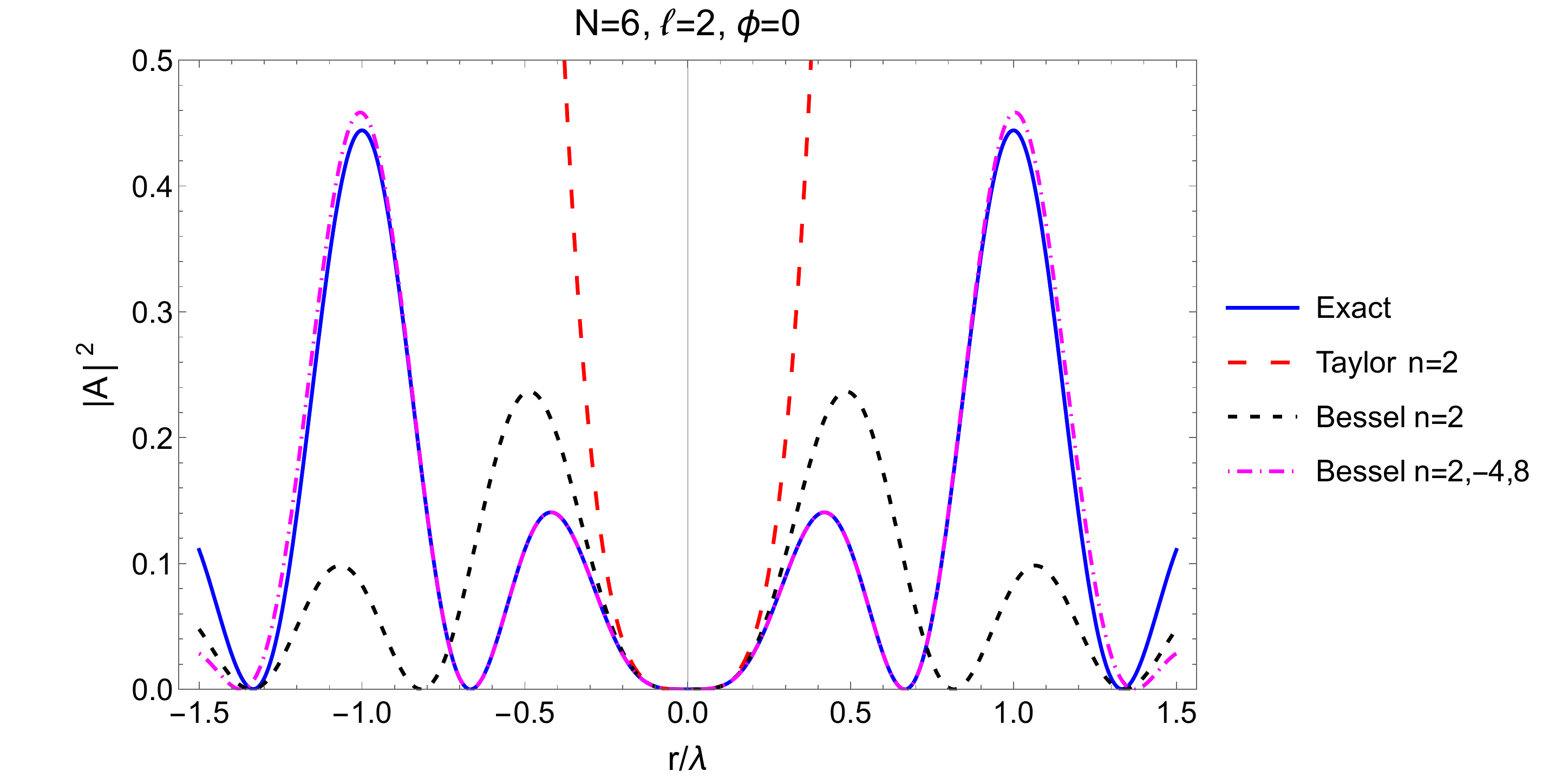}
  \includegraphics[width=13.cm]{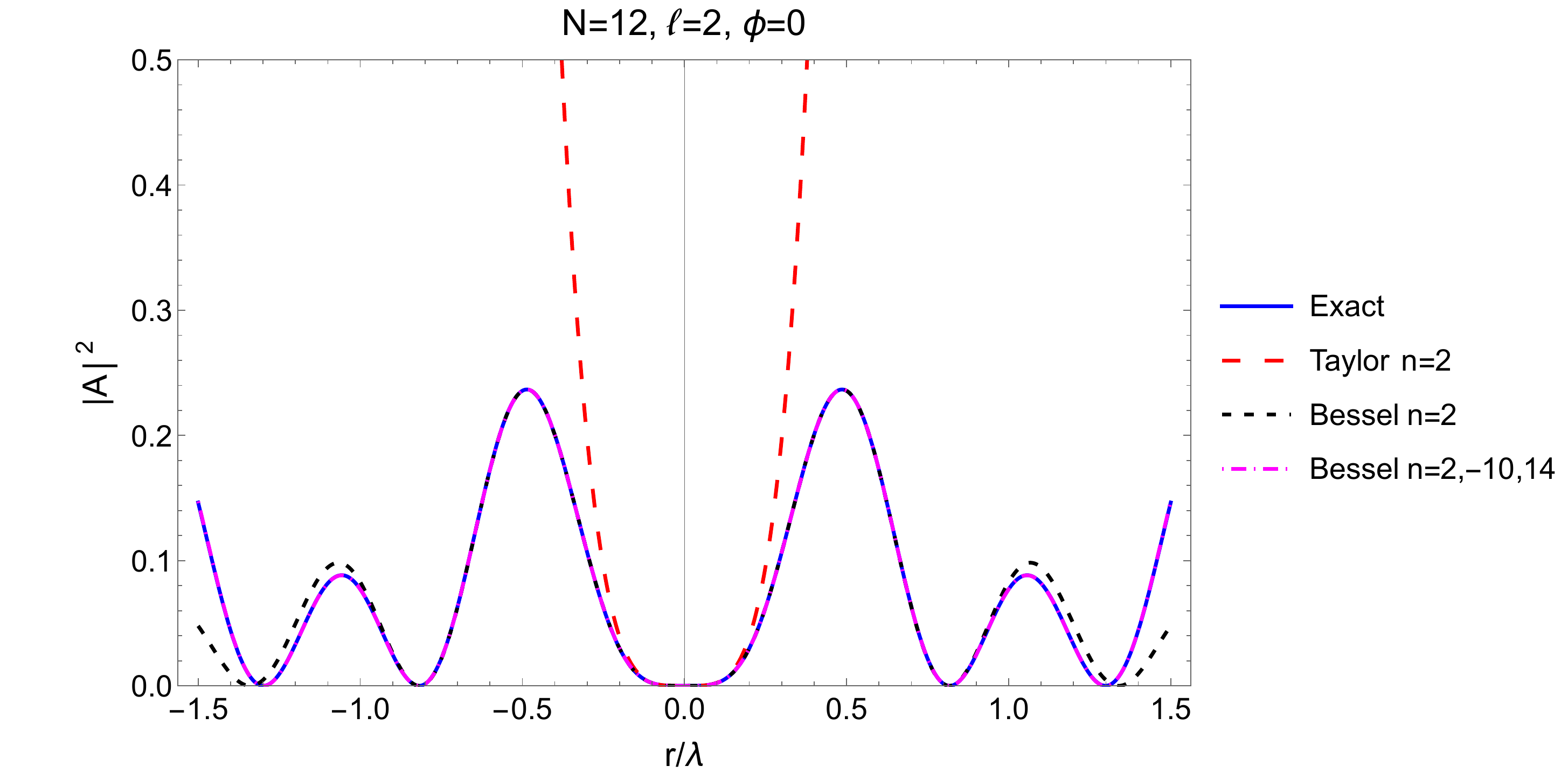}
\caption{Same as Figure 2 for an input phase parameter $l=2$. Notice $(kr)^2$ behavior for (a) and $(kr)^4$ for (b), (c) as $(kr)\to 0$. See text for explanation.}
\end{figure}

\begin{figure}[htbp]
\label{figarray3}
  \centering
  \includegraphics[width=7.cm]{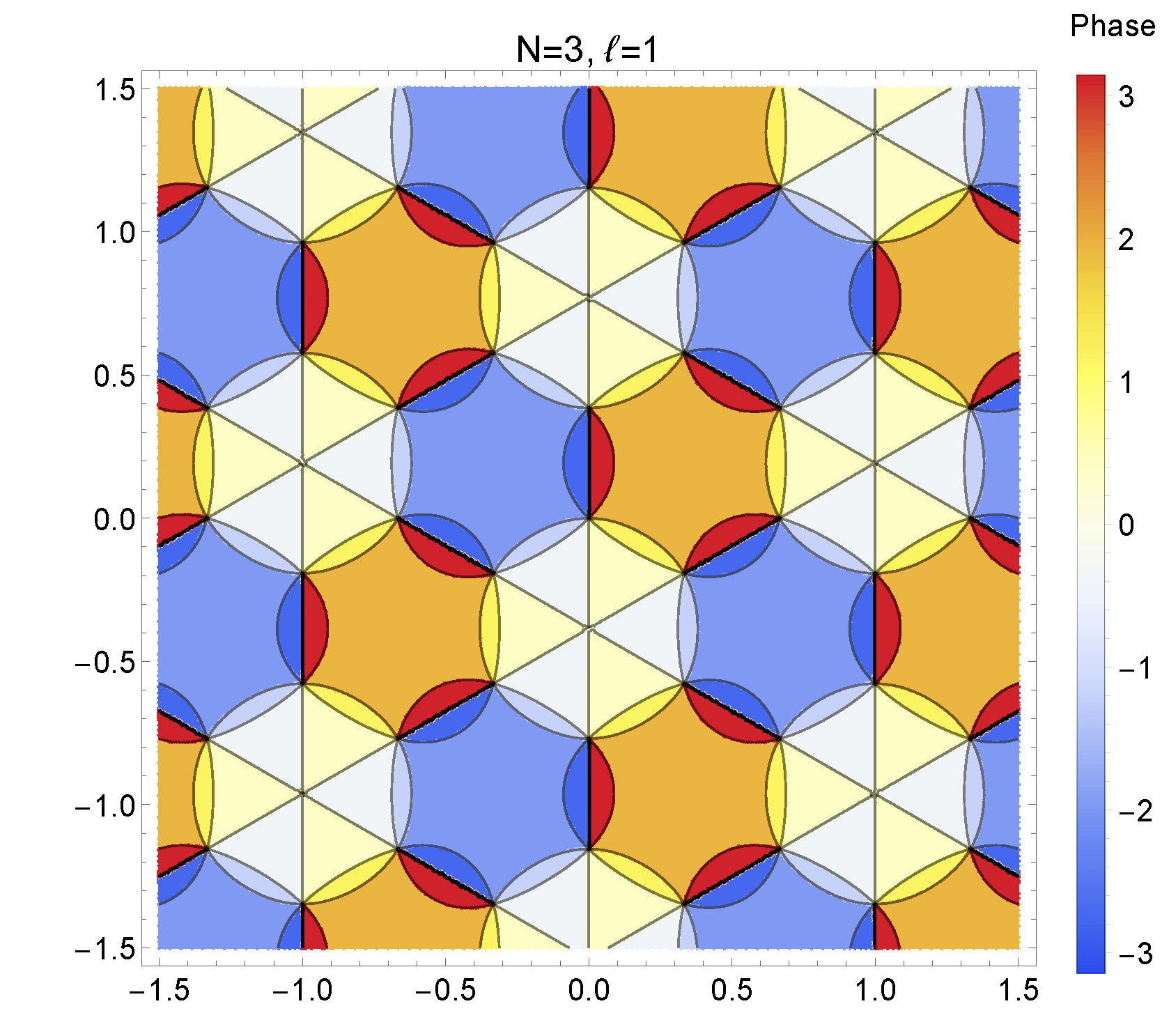}
  \includegraphics[width=7.cm]{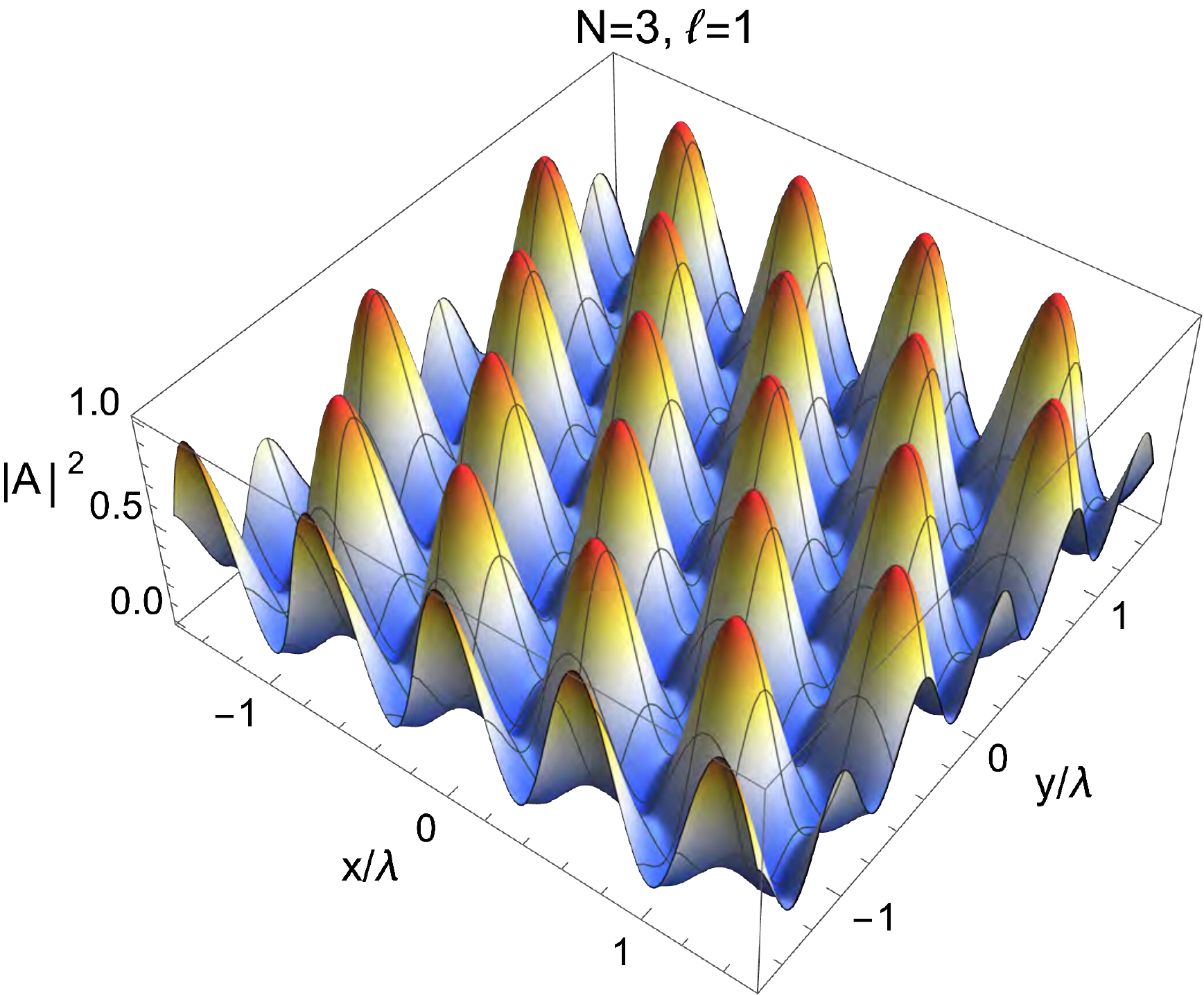}
  \includegraphics[width=7.cm]{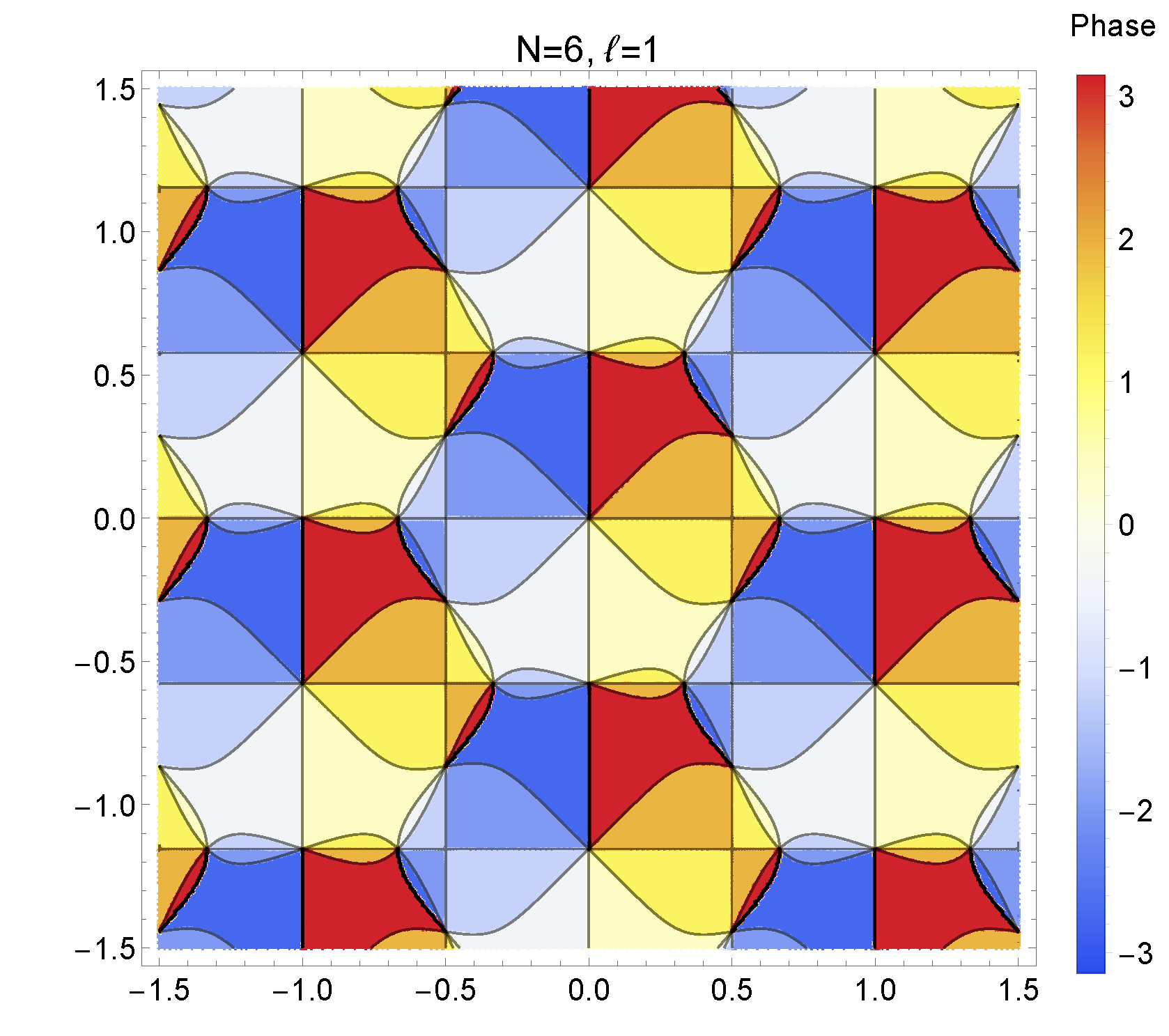}
  \includegraphics[width=7.cm]{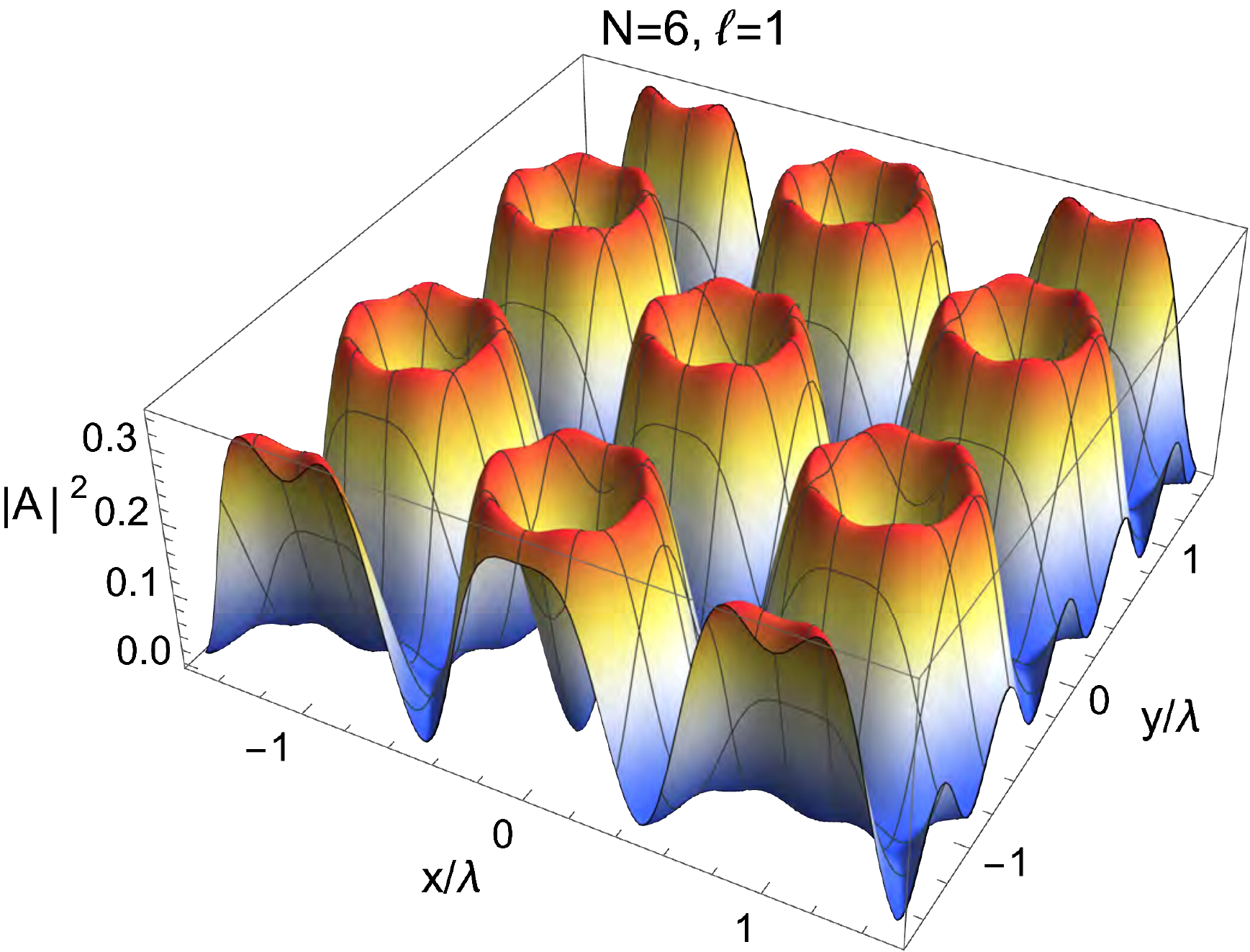}
  \includegraphics[width=7.cm]{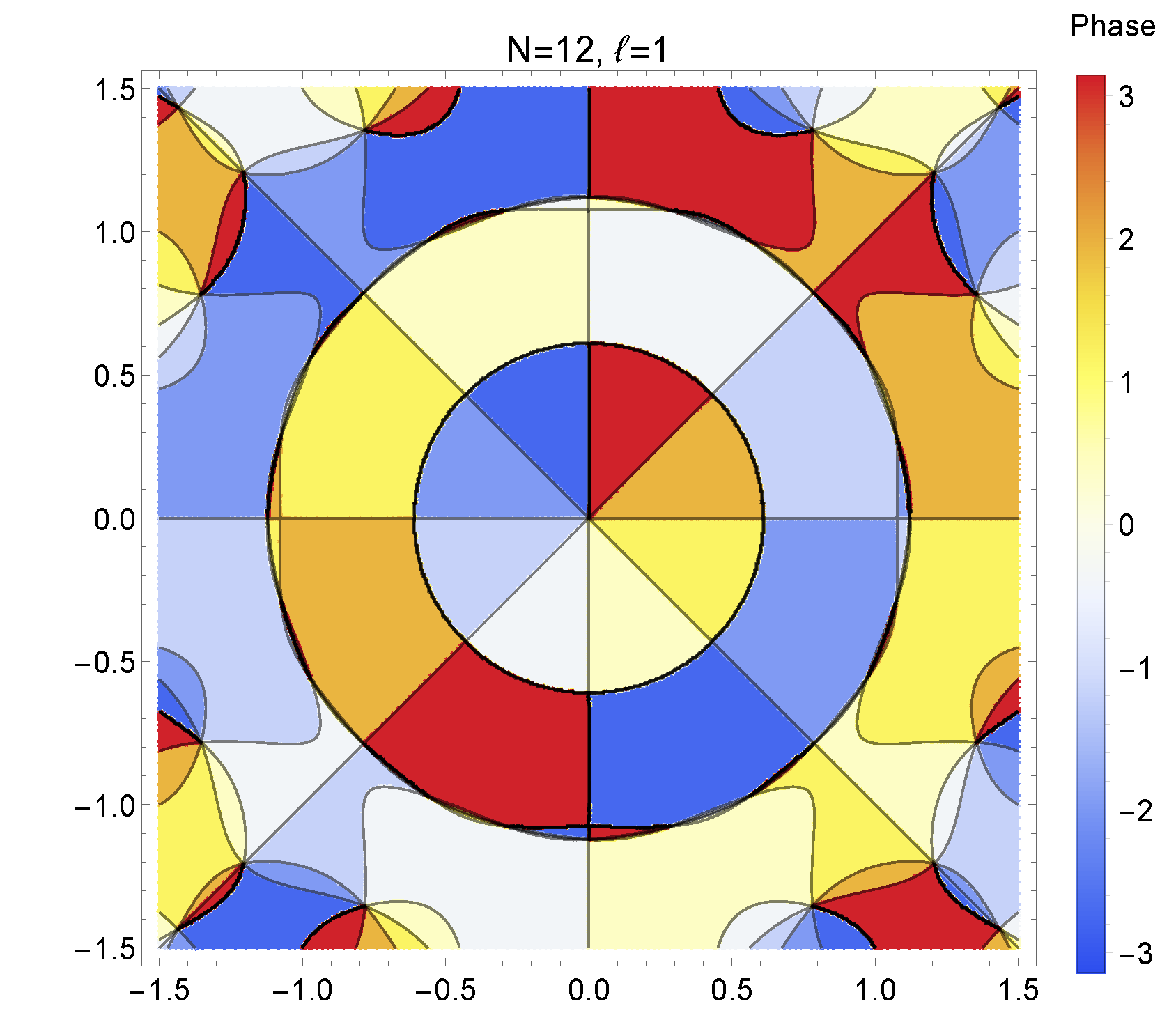}
  \includegraphics[width=7.cm]{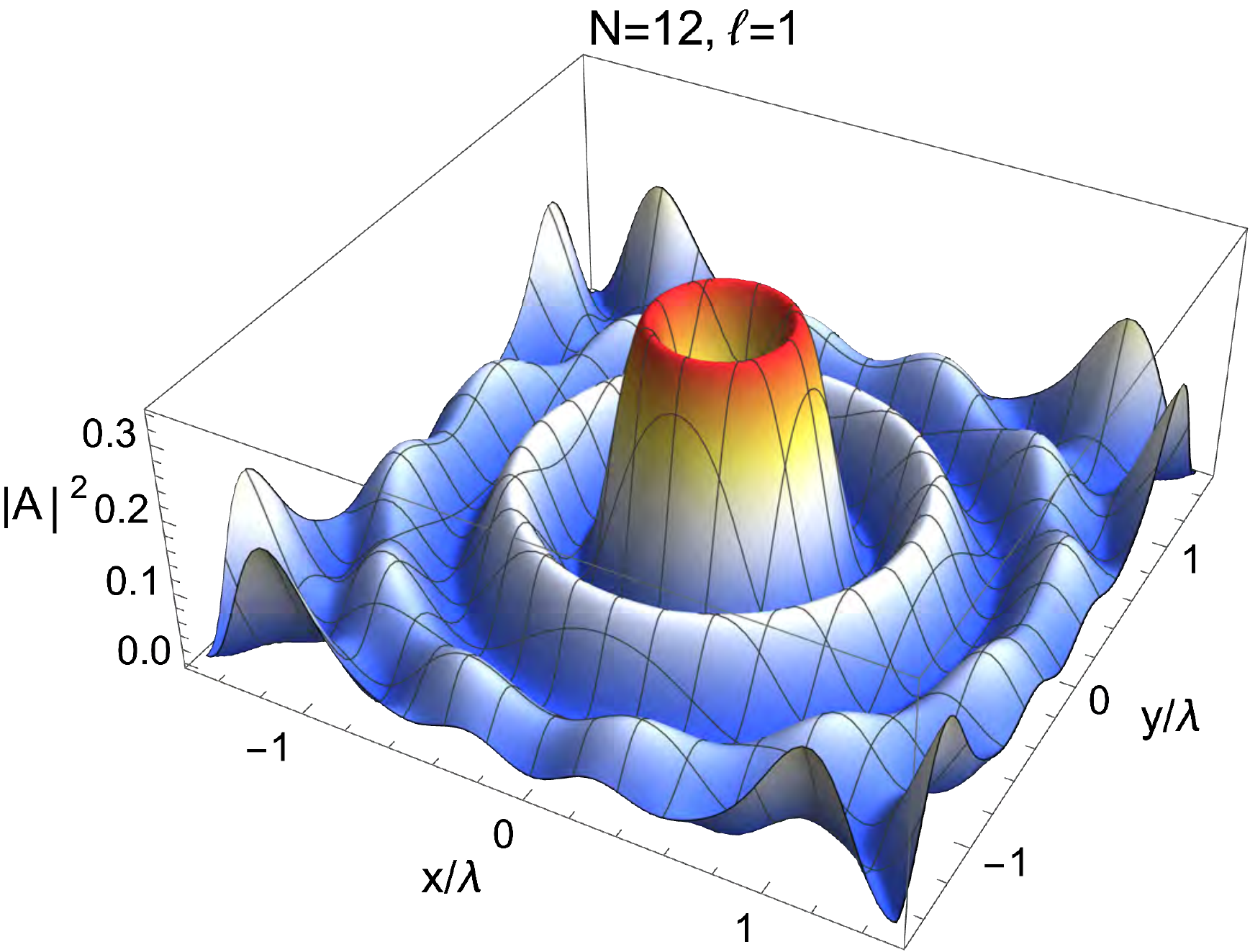}
\caption{Phases, in radian (left column) and intensities, in arbitrary units (right column) for wave superposition Eq.(\ref{eq:Amps}) with $l=1$ and, top to bottom, $N=$ 3, 6 and 12. The change from the lattice of vortices to a single vortex with increasing $N$ is due to cancellation of lowest-order non-leading Bessel terms as specified in Table 2. }
\end{figure}

The results for $l=1$ are shown in Fig.~2 for the time-period-averaged amplitude squared (or intensity) at a fixed polar angle $\phi=0$ ($r>0$) and $\phi=\pi$ ($r<0$). It can be seen that the result quickly approaches $N\to\infty$ limit even for a dozen elements, $N=12$. Shown in Fig.~2 are contributions from leading order Taylor terms (see Table~1), a leading order Bessel term ($n=1$) and the effect from including two next-to-leading order Bessel vortices in accordance with the expansion of Eq.(\ref{eq:2Dsum}). Note that for each element number $N$, orders of additional terms are different. Namely, for $l=1$ we have $n=1, -2,4$ ($N=3)$, $n=1,-5,7$ ($N=6$) and $n=1,-11,13$ ($N=12$), as follows from $n=l+mN$ condition (see also Table 2). 

To generate a vortex with a topological charge $n=2$, we apply an input phase parameter $l=2$, with the resulting intensity shown in Fig.~3. It can be seen that the result also quickly approaches a single-Bessel vortex limit $n=2$ even for a dozen elements, $N=12$, but for lower $N$ the vortex structure is stronger affected by next-to-leading terms. Applying $l=2$ in the $N=3$ case leads to a lower-order leading vortex $n=-1$ that has a sign opposite to $l$, which follows from $n=l+mN$ relation for $(l,N,m)=(2,3,-1)$. The following states are included for the vortex expansion Eq.(\ref{eq:2Dsum}) with $l=2$ are: $n=-1, 2,-4$ ($N=3)$, $n=2,-4,8$ ($N=6$) and $n=2,-10,14$ ($N=12$).

The phase of the amplitude $arg(A(\vec r,t=0)$ and related intensities are shown in Fig.~4 for $l=1$. The time dependence is defined by an overall factor $\exp[-i\omega t]$. One can observe a qualitative change with increasing element number $N$. For all values of $N\geq 2n+1=3$, we can clearly identify the central point that has a phase singularity, $i.e$ the amplitude for $r\to 0$ is $\propto r e^{i\phi}$ and its phase for $r=0$ is undefined. However, we do not see a doughnut structure of the vortex for $N=3$. For higher values of $N$,  we obtain an array of doughnuts for $N=6$ and we obtain a single-doughnut structure for $N=12$. Also, for lower $N$ secondary vortices appear (with the same topological charge as the central vortex), but these secondary vortices are eliminated by increasing the element number $N$. Such behavior can be attributed to higher-order Bessel states in Eq.(\ref{eq:2Dsum}), as also revealed in Fig.~2. 

In a previous study, the authors of \cite{Yang13} performed a numerical study of the problem for acoustical waves and concluded that to produce vortices with a maximum topological charge $l$, the minimum source number of $N_{min}=max(2|l|+1,4)$ is required, thereby ruling out an $N=3$ option for $l=1$ vortex. Here we provide analytic justification of the $N_{min}=2|l|+1$ requirement, with $N=3$ included.

\section{Summary}

We considered the generation of  vortex wave states from coherent superpositions of waves from discrete and individually-phased elements arranged in a circular planar array, paying special attention to sparse arrays with minimal element numbers. Using both Taylor and Jacobi-Anger expansion of individual waves, we analyzed vortex topologies for different element numbers  $N$ and different values of an input phase parameter $l$. For a given value of $l$, a plurality of vortices may be generated which topological charges depend both on $l$ and on $N$. For a circular array of $N$ elements and a fixed parameter $l$, the resulting vortex is an equal-weight  superposition of Bessel vortex states with allowed topological charges $n=l+mN$, where $m$ is an integer.  This sets a constraint on the element number $N$ necessary to generate a desired topological charge $n$ of the lowest-order vortex, namely, $N\ge2n+1$. 
If this condition is not met, lower-order vortices with signs opposite to $l$ would be generated. As $N$ increases, the Bessel vortex orders adjacent to $n=l$ are eliminated -- in accordance with $n=l+mN$ -- and for $N\to\infty$ only a single vortex with $n=l$ remains, recovering a single-order Bessel-vortex result known from the literature Ref.~\cite{Courtney13}.
If the element number is twice the input phase parameter, $N=2l$, a superposition of vortices with opposite topological charges, $(\ket{l}+\ket{-l}+\ket{3l}+\ket{-3l}+...)$, is formed.
Such states may have potential applications as qubits for quantum computing or quantum communications. Another possible application is for material characterization, since due to vortex dichroism \cite{Forbes19} the states with opposite topological charges may evolve differently as they propagate in matter.

Our results provide practical limitations on the generation of vortices from sparse arrays of wave sources. For example, the choice of eight elements (N=8) for plasmonic nano-antennas Refs.\cite{Arikawa17,Shutova20} makes it possible  to generate a state with a leading-order vortex up to $l$=3; while a four-element RF antenna (N=4) considered in Ref.\cite{Mika20} would yield at most $l$=1.

Extension of the presented approach to formation of optical vortices from a coherent superposition of $N$ laser beams is currently in progress.

\ackn{
We thank US Army Research Office for support under Grant W911NF-19-1-0022. In addition, KS acknowledges undergraduate student apprenticeship support from AEOP Program of US Army Research Office in the summer 2021. We appreciate help from Norbert Linke for reading the manuscript and for useful comments.}

\end{document}